\documentclass{amsart}
\newtheorem{theo}{Theorem}[section]
\newtheorem{lemm}[theo]{Lemma}
\newtheorem{exem}[theo]{Example}
\newtheorem{rema}[theo]{Remark}
\newtheorem{defi}[theo]{Definition}
\newtheorem{propo}[theo]{Proposition}

\def\lcm{\rm lcm}
\def\N{\mathbb{N}}
\def\F{\mathbb{F}}

\begin{document}

\title[Second Weight]{The Second Weight of Generalized Reed-Muller Codes in Most Cases}

\author{Robert Rolland}

\address{Institut de Math\'ematiques de Luminy\\
Case 907, 13288 Marseille cedex 9, France\\
http://robert.rolland.acrypta.com}
\email{robert.rolland@acrypta.fr}
\date{\today}

\begin{abstract}
The second weight of the Generalized Reed-Muller code of length
$q^n$ and
order $d$ over the finite field with $q$ elements
is now known for $d <q$ and $d>(n-1)(q-1)$. In this paper, we determine
the second weight for the other values of $d$ which are not
 multiples of
$q-1$ plus $1$. For the special case $d=a(q-1)+1$ we give an estimate.
\end{abstract}

\maketitle

\section{Introduction - Notations}
\label{sec:un}
Let $\F_q$ be the finite field with $q$ elements 
and $n \geq 1$ an integer.
Let $d$ be an integer such that
$1\leq d < n (q-1).$
The generalized Reed-Muller code of order $d$ 
is the following subspace of the space $\F_q^{(q^n)}$:
$${\rm RM}_q(d,n)=\left\{\bigl( f(x) \bigr)_{x\in \F_q^n}
~| ~ f \in \F_q[X_1,\ldots,X_n] \hbox{ and } {\deg}(f) \leq d\right\}.$$
It may be remarked that the polynomials $f$ determining this code are viewed 
as polynomial functions. Hence each codeword is associated with a unique 
reduced polynomial, namely a polynomial  
whose partial degrees are $\leq q-1$.
We will denote by ${\mathcal F}(q,d,n)$ the space of the reduced polynomials $f$
$\in \F_q[X_1,\ldots,X_n]$ such that $\deg  f \leq d$.
From a geometric point of view a polynomial $f$
defines a hypersurface in $\F_q^n$ and the number of points $N(f)$
of this hypersurface (the number of zeros of $f$) is related to the weight
of the associated codeword by the following formula:
$$W(f)=q^n-N(f).$$
The code ${\rm RM}_q(d,n)$ has the following parameters:
\begin{enumerate}
 \item length  $m=q^n$,
 \item dimension  $k=\sum_{t=0}^d\sum_{j=0}^n (-1)^j 
\left( \begin{array}{c} n\\j \end{array} \right) 
\left( \begin{array}{c} t-jq+n-1\\t-jq \end{array} \right)$,
 \item minimum distance $ W_1=(q-b) q^{n-a-1},$
where $a$ and $b$ are the quotient and the remainder
in the Euclidian division of $d$ by $q-1$, namely
$d=a (q-1)+b$ and $0 \leq b <q-1$.
\end{enumerate}

\begin{rema}
 Be carefull not to confuse symbols. With our notations,
the Reed-Muller code of order $d$ has length $m$, dimemsion $k$
and minimum distance $W_1$. Namely it is a $[m,k,W_1]-$code.
The integer $n$ is the number of variables of the polynomials
defining the words and the order $d$ is the
maximum total degree of these polynomials.
\end{rema}

The minimum distance was given by T. Kasami, S. Lin, W. Peterson in \cite{KLP}.
The words reaching this bound were characterized by P. Delsarte, J. Goethals
and F. MacWilliams in \cite{DGMW}.
Let us denote by $W_2$, the second weight, namely the weight
just above the minimum distance. 
If $d=1$, we know that the code has only three weights: $0$, the minimum distance 
$W_1=q^n-q^{n-1}$ and the second weight $W_2=q^n$.
For $d=2$ and $q=2$ the weight distribution is  more or less a
consequence of the investigation of quadratic forms done by L. Dickson
in \cite{dick} and was also done by E. Berlekamp and N. Sloane in an 
unpublished paper. For $d=2$ and any $q$ (including $q=2$) 
the weight distribution was given
by R. McEliece
in \cite{McE}. For $q=2$, for any $n$ and any $d$, the weight distribution is 
known in the range $[W_1,2.5W_1]$ by a result of Kasami, Tokora, Azumi \cite{kta}. 
In particular, the second weight is $W_2=3\times 2^{n-d-1}$.
For $d \geq n (q-1)$ the code ${\rm RM}_q(d,n)$ is the whole 
${\mathcal F}(q,d,n)$,
hence any integer $0\leq t \leq q^m$ is a weight. 
The second weight was first studied by J.-P. Cherdieu and R. Rolland in \cite{CR}
who proved that when $q>2$ is fixed, for $d<q$ sufficiently small
the second weight is
$$W_2=q^n - d q^{n-1} - (d-1) q^{n-2}.$$
Their result was improved by A. Sboui 
in \cite{AS}, who
proved the formula for $d \leq q/2$. The methods in \cite{CR}
and \cite{AS} are of a geometric nature by means of which the
codewords reaching this weight can be determined. These codewords
are hyperplane arrangements.
Recently, O. Geil in \cite{Geil}, using Gr\"obner basis
methods, proved the formula for $d <q$. 
Moreover as an application of his method, he gave a new
proof of the Kasami-Lin-Peterson minimum distance formula and determined, when
 $d>(n-1) (q-1)$,
the first $d+1 -(n-1) (q-1)$ weights.
However
the Gr\"obner basis
method does not determine all the codewords reaching the second weight.

To summarize the state of the art, let us note the following main points

\begin{enumerate}
 \item for $q=2$, the second weight is known;
 \item for $n=2$, the second weight is known for all values of $d$;
 \item for $n>2$, the second weight is known for $d<q$ and for $d>(n-1) (q-1)$.
\end{enumerate}

Here and subsequently,
$a$ and $b$ are respectively the quotient and the remainder
in the Euclidian division of $d$
by $q-1$.
In this paper, we determine for $n \geq 3$, $q \geq 3$ and 
$b \neq 1$
the second weight $W_2$ (or the second
number of points of a hypersurface $N_2=q^n-W_2$)
of the generalized Reed-Muller code and for $b=1$ we give 
a lower bound on this second weight. This work is done
for all the other values of $d$ not yet handled,
namely  
$q \leq d \leq (n-1) (q-1) \hbox{ for } q \geq 3.$ 
Let us remark that for such a $d$, we have
$1 \leq a \leq (n-1).$
Moreover, if $a=(n-1)$ then $b=0$. 
If $f \in {\mathcal F}(q,d,n)\setminus \{0\}$ we will denote by 
$N(f)$ the number of zeros of $f$ {\it i.e.} the number of points of
the hypersurface defined by $f$, and by $W(f)=q^n-N(f)$ the weight
of the associated codeword. If $f=h_1 h_2 \ldots h_d$ where
$h_i(X_1,\ldots ,X_n)$ is a polynomial of degree $1$, we consider
the hyperplane arrangement ${\mathcal A}=\{H_i\}_{i=1,\ldots,d}$
where $H_i$ is the affine hyperplane defined by $h_i(X_1,\ldots,X_n)=0$.
The hypersurface defined by $f$ is the union of the hyperplanes $H_i$.
We will set
$$N({\mathcal A})=N(f)=\# \cup_{i=1}^d H_i, 
\quad W({\mathcal A})=q^n-
N({\mathcal A}).$$

The paper is organized as follows. We begin in Section \ref{sec:deux} with
a result on some special hypersurfaces: those which are
unions of affine hyperplanes defined by linearly independant
linear forms. We determine the configurations of this class
having the minimal weight among those which do not reach the 
minimum distance ({\it i.e.} which are not maximal). 
It turns out that these particular hypersurfaces reach
the second weight except possibly for the case $d=a(q-1)+1$.
In Section \ref{sec:trois}  we state and prove the main theorem
on the value of the second weight for general hypersurfaces.
The proof which follows the method introduced
by O. Geil in \cite{Geil} is based on Gr\"obner basis techniques. 
It also uses a tedious combinatorial lemma whose proof is 
done in the appendix.
We point out in Section \ref{sec:quatre} some open questions 
related to the case $d=a(q-1)+1$
not solved in this paper and to the determination of the codewords
reaching the second weight.

\section{Blocks of hyperplane arrangements}\label{sec:deux}
\subsection{Basic facts}
Let us suppose that $d=d_1+d_2+\ldots+d_k$ where
$$
\left \{
\begin{array}{lllll}
1 &\leq &d_i &\leq & q-1, \\
1 & \leq & k &\leq & n.
\end{array}
\right .
$$ 
Let us denote by $f_1,f_2, \ldots,f_k$, $k$ independant
linear forms on $E=\F_q^n$, and let us consider the following
hyperplane arrangement:
for each $f_i$ we have $d_i$ distinct parallel hyperplanes
defined by
$$f_i(x)=u_{i,j} \quad 1 \leq j \leq d_i.$$
This arrangement of $d$ hyperplanes is
consists of $k$ blocks of parallel hyperplanes, the $k$ directions of the blocks
being linearly independant. 
The set of such hyperplane arrangements will be called ${\mathcal L}$.
\begin{theo}\label{main}
Let ${\mathcal A}$ be a hyperplane arrangement in ${\mathcal L}$
and let us set
$$A=\bigcup_{H\in {\mathcal A}} H.$$
Then, the number of points of $A$ is
$$N({\mathcal A})=\# A=q^n - q^{n-k}  \prod_{i=1}^k (q-d_i).$$
\end{theo}
\noindent {\bf Proof.}  We can suppose that $f_i(x)=x_i$. The points which are not in $A$
satisfy the following conditions: 
$$
\begin{array}{cccc}
 &x_1 &\neq & u_{1,1}, u_{1,2},\ldots,u_{1,d_1},\\
\hbox{and} & & & \\
 &x_2 & \neq & u_{2,1}, u_{2,2},\ldots,u_{2,d_2},\\
\hbox{and} & & & \\
 & \vdots & &  \\
\hbox{and} & & & \\
 &x_k & \neq & u_{k,1}, u_{k,2},\ldots,u_{k,d_k}.
\end{array}
$$
Moreover for $u>k$, the $x_u$ are arbitrary.
Hence the number of points which are not in $A$ is
$$q^{n-k}  \prod _{i=1}^k (q-d_i).$$
\begin{exem} Let $k=a+1$, $d_i=q-1$ for $i=1,2,\ldots,a$ and $d_{a+1}=b$.
We know that these configurations are the maximal configurations,
namely the configurations ${\mathcal A}$ such that
$N({\mathcal A}) =q^n-W_1=N_1$. 
\end{exem}
\begin{rema}
The number $N({\mathcal A})$ depends only on $k$ and $d_1,d_2,\ldots,d_k$.
These values define a type $T$ (i.e. the set of all arrangements 
in ${\mathcal L}$
with the same values $k$ and $d_1,d_2,\ldots,d_k$). 
We will denote by $N(T)$ the common number of points of 
all the type $T$ arrangements .
\end{rema}

\subsection{Modification of a maximal configuration when $q \geq 3$}
Let us start from  a maximal configuration ${\mathcal A}$,
then
$$N({\mathcal A})=\# \bigcup_{H \in {\mathcal A}} H=N_1.$$
We know (cf. \cite{DGMW}) that a maximal configuration is given 
by $a+1$ linearly independant
linear forms $f_1f_2,\ldots f_{a+1}$ such that the $d=a (q-1)+b$ hyperplanes
are constituted by the following blocks:
\begin{enumerate}
 \item {\bf $a$ blocks of $q-1$ parallel hyperplanes:}\\ 
for each $i \in \{1,\ldots,a\}$ let $A_i=\{u_{i,j}\}_{1\leq j\leq q-1}$
be a subset of $\F_q$ such that $\# A_i=q-1$.
We denote by ${\mathcal A}_i$ the block of the  
$q-1$ distinct parallel hyperplanes $H_{i,j}$ defined by
$$H_{i,j}=\{x \in E~|~f_i(x)=u_{i,j}\};$$
 \item {\bf one block of $b$ parallel hyperplanes:}\\
let $B=\{v_j\}_{1\leq j\leq b}$ be a subset of $\F_q$ such that 
$\# B=b$. We denote by ${\mathcal B}$ the bloc of $b$ distinct parallel
hyperplanes $P_j$ defined by
$$P_j=\{x \in E~|~f_{a+1}(x)=v_j\}.$$
Let us remark that if $b=0$, then $B=\emptyset$ and the block
${\mathcal B}$ is void.
\end{enumerate}
A maximal configuration is in ${\mathcal L}$.

\subsubsection{Type $1$ exchange}
The type $1$ exchange replaces one hyperplane of a complete block
by a hyperplane in the last block. The so obtained configuration 
is in ${\mathcal L}$
and is not maximal by the characterization of P. Delsarte, J. Goethals
and F. MacWilliams.

More precisely, we suppose that 
$1 \leq a \leq n-1$
and 
$0 \leq b < q-2.$
(For $b=q-2$ this exchange gives  another  maximal arrangement.)
Let us define the following transform of the configuration ${\mathcal A}$.
Choose $i \in \{1,\ldots,a\}$, $j \in \{1, \ldots,q-1\}$ and 
$v_{b+1} \in \F_q \setminus B$. Replace the hyperplane
$H_{i,j}$ by the hyperplane $P_{b+1}=\{x \in E~|~f_{a+1}(x)=v_{b+1}\}$.
We call $T_1$ the type of the obtained configuration. 
\begin{propo} \label{itb}
For $1 \leq a \leq n-1$ and $0 \leq b < q-2$,
the following formulas hold:
$$N({T_1})=q^n-2 q^{n-a-1}  (q-b-1),$$
$$N_1-N({T_1})=q^{(n-a-1)} (q-b-2)>0.$$
\end{propo}
\noindent {\bf Proof.} 
The first formula is a direct consequence of Theorem \ref{main}.
A direct computation gives us the second formula.

\subsubsection{Type $2$ exchange}
The type $2$ exchange replaces one hyperplane of a complete block
by a hyperplane defined by a new linear form, linearly independant from the $a+1$ original ones. The obtained configuration is in ${\mathcal L}$ and is not maximal.

More precisely, we suppose that 
$1 \leq a <n-1$
 and 
$1 \leq b <q-1.$
(for $a=n-1$ the type $2$ exchange cannot be done, and for $b=0$
it is the type $1$ exchange).
Choose a linear form $f_{a+2}$ that together with
the linear forms $$f_1,\ldots,f_{a+1},f_{a+2}$$ 
forms a linearly independent system.
Choose $i \in \{1,\ldots,a\}$, $j \in \{1, \ldots,q-1\}$, 
$w \in \F_q$ and 
replace the hyperplane
$H_{i,j}$ by the hyperplane $Q=\{x \in E~|~f_{a+2}(x)=w\}.$
We call $T_2$ the type of the new obtained arrangement. 
\begin{propo}
For $1 \leq a <n-1$ and $1 \leq b <q-1$
the following formulas hold:
$$N({T_2})=q^n-2 q^{n-a-2}  (q-1) (q-b),$$
$$N_1-N({T_2})= q^{(n-a-2)} (q-b) (q-2)>0.$$
\end{propo}
\noindent {\bf Proof.} 
The first formula is a direct consequence of Theorem \ref{main}.
A direct computation gives the second formula.

Now let us compare $N({T_1})$ and $N({T_2})$ for $d$ such that
$$1 \leq a \leq n-2,$$
$$1 \leq b <q-2.$$
A simple computation gives the following:
\begin{propo}
For $1 \leq a \leq n-2$ and $1 \leq b <q-2$, we get
$$N({T_1})-N({T_2})=2 q^{(n-a-2)} b>0.$$
\end{propo}

\subsubsection{Type $3$ exchange}
The type $3$ exchange replaces one hyperplane of the last block
by a hyperplane defined by a new linear form, linearly independant from the 
$a+1$ original ones. The obtained configuration is in ${\mathcal L}$ and
is not maximal.

We suppose that 
$1 \leq a <n-1$ 
and
$2 \leq b <q-1.$
(For $b=1$, the exchange does not change the type of the configuration).
Choose a linear form $f_{a+2}$ which constitutes
with the linear forms $f_1,\ldots,f_{a+1}$ a linearly independant system.
Choose $j \in \{1,\ldots,b\}$ and 
$w \in \F_q$. Replace the hyperplane
$P_j$ by the hyperplane $Q=\{x \in E~|~f_{a+2}(x)=w\}$.
We call $T_3$ the type of the new obtained arrangement. 
\begin{propo}
For $1 \leq a <n-1$ and $2 \leq b <q-1$
the following formulas hold:
$$N({T_3})=q^n-q^{n-a-2} (q-1) (q-b+1),$$
$$N_1-N({T_3})= q^{(n-a-2)} (b-1)>0.$$
\end{propo}
\noindent {\bf Proof.} 
The first formula is a direct consequence of Theorem \ref{main}.
A direct computation gives the second formula.

Now let us compare
$N({T_1})$ and $N({T_3})$ for $d$ such that
$1 \leq a \leq n-2,$
$2 \leq b <q-2.$
A simple computation gives the following:
\begin{propo}
 For $1 \leq a \leq n-2$ and $2 \leq b <q-2$, we get
$$N({T_3})-N({T_1})=q^{(n-a-2)} (q^2-(b+2)q -b+1) > 0.$$
\end{propo}

For $b=q-2$, the type $1$ transform is not valuable (it gives $N({T_1})=N_1)$
so we must compare $N({T_3})$ and $N({T_2})$.
A direct computation gives the following:
\begin{propo}
 For $1 \leq a \leq n-2$ and $b=q-2$, 
$$N({T_3})-N({T_2})=q^{(n-a-2)}(q-1) >0$$
holds.
\end{propo}

\subsubsection{Type $4$ exchange}
The type $4$ exchange, used when $b=1$, 
deletes the unique hyperplane of the last block.
Let us denote
by $T_4$ the type of the new obtained arrangement. Let us remark that
this configuration is the maximal configuration related to
the degree $d-1$, namely gives the minimal distance for the Reed-Muller code of order $d-1$.
Then by a direct computation the following proposition holds:
\begin{propo}
For $1 \leq a <n-1$ and $b=1$
the following formulas hold:
$$N({T_4})=q^n-q^{n-a},$$
$$N_1-N({T_4})= q^{(n-a-1)} >0.$$
\end{propo}

Now let us compare, for $b=1$ and $q=3$,
$N({T_2})$ and $N({T_4}$). A simple computation
gives the following:
\begin{propo}
 For $q=3$, $1 \leq a \leq n-2$ and $b=1$, we get
$$N(T_2)-N(T_4)=3^{n-a-2}.$$
\end{propo}
Let us also compare, for $b=1$ and $q \geq 4$,
$N({T_1})$ and $N({T_4}$). 
A simple computation gives the following:
\begin{propo}
 For $q\geq 4$, $1 \leq a \leq n-2$ and $b=1$, we get
$$N({T_4})-N({T_1})=q^{n-a-1}(q-4)\geq 0.$$
\end{propo}

\subsubsection{The best case for a type $T_1$ or $T_2$ or $T_3$ or $T_4$ arrangement}
Let us set $N'_2=\max(N(T_1),N(T_2),N(T_3),N(T_4))$
(if $N(T_i)$ is not defined we don't consider it in the $\max$).
$N'_2$ is the largest number of zeros for a type $T_1$ or $T_2$ or $T_3$ or
$T_4$ arrangement. 
We summarize the results of this subsection in
the following theorem.
We will denote by $W'_2$ the second weight for the arrangements 
of the previous type,
namely $W'_2=q^n- N'_2$.
\begin{theo}
The values of $N'_2$ and $W'_2$ are:
\begin{enumerate}
\item Let us suppose that $q \geq 4$.
  \begin{enumerate}
     \item For 
$1 \leq a <n-1
\hbox{ and }
2\leq b <q-1,$ the maximal number of points
$N'_2$ is reached by the type 
$T_3$, hence
$$N'_2=N({T_3})=q^n - q^{n-a-2} (q-1) (q-b+1),$$
$$W'_2=q^{n-a-2} (q-1) (q-b+1).$$
     \item For 
$1 \leq a <n-1 \hbox{ and }
b=1,$ the maximal number of points
$N'_2$ is reached by the type 
$T_4$, hence
$$N'_2=N({T_4})=q^n - q^{n-a},$$
$$W'_2=q^{n-a}.$$
     \item For
$1 \leq a\leq n-1 \hbox{ and } b=0,$ the maximal number of points
$N'_2$ is reached by the type
$T_1$, hence
$$N'_2=N({T_1})=q^n - 2 q^{n-a-1} (q-b-1),$$
$$W'_2=2 q^{n-a-1} (q-b-1).$$
  \end{enumerate}
\item Let us now suppose that $q=3$.
\begin{enumerate}
  \item For 
$1 \leq a \leq n-1 \hbox{ and } b=0,$ the maximal number of points
$N'_2$ is reached by the type $T_1$, hence
$$N'_2=N({T_1})=q^n - 2 q^{n-a-1} (q-1),$$
$$W'_2=4 \times 3^{n-a-1}.$$
   \item For 
$1 \leq a < n-1 \hbox{ and } b=1,$ the maximal number of points
$N'_2$ is reached by the type
$T_2$, hence 
$$N'_2=N({T_2})=q^n - 2 q^{n-a-2} (q-1)^2,$$
$$W'_2=8 \times 3^{n-a-2}.$$
  \end{enumerate}
\end{enumerate}
\end{theo}

\subsection{The best case for a ${\mathcal L}$ arrangement}
\begin{theo}\label{th:hyper}
Let ${\mathcal B}$ a hyperplane arrangement in ${\mathcal L}$. Suppose
that ${\mathcal B}$ is not maximal and not in $T_1$ nor in $T_2$ nor in $T_3$
nor in $T_4$.
Then $N({\mathcal B})< N'_2$.
\end{theo}
 \noindent {\bf Proof.} 
Let us denote by $k$, $d_1,\ldots,d_k$ the values
defining the type of this arrangement. 
Then $$N({\mathcal B})=q^n - q^{n-k}  \prod_{i=1}^k (q-d_i).$$
Let us set $d'=\sum_{i=1}^k d_i=a'(q-1)+b'$.

\begin{enumerate}
\item If we can find two distinct indices $i_1$ and
$i_2$ such that 
$$1 \leq d_{i_1} \leq d_{i_2} \leq q-2,$$ 
let us replace one hyperplane of the block $i_1$
by a new hyperplane (not in ${\mathcal B}$) added to the block $i_2$.
We obtain the arrangement ${\mathcal B}'$.
As ${\mathcal B}$ is not in 
$T_1$ nor in $T_2$ nor in $T_3$, ${\mathcal B}'$ is not a maximal arrangement.
Moreover
$$N({\mathcal B}')-N({\mathcal B})=
K((q-d_{i_1}) (q-d_{i_2})-(q-d_{i_1}+1) (q-d_{i_2}-1))$$
$$=K(d_{i_2}-d_{i_1}+1)>0$$
where $K=q^{n-k} \prod_{i\neq i_1,i_2} (q-d_i)$.
Then ${\mathcal B}$ is not maximal among the ${\mathcal L}$ arrangements
not reaching $N_1$.
\item If all the $d_i$ but $d_{i_1}$ are $0$ or $q-1$, namely
${\mathcal B}$ consists of $a'$ complete blocks
containing $q-1$ hyperplanes and one block of $b'$ hyperplanes.
As ${\mathcal B}$ is not a maximal configuration
then either
$a'<a$ holds or $a'=a$ and $b'<b$ holds.
In both cases $d'=a'(q-1)+b'<d$ and we can add a new direction, linearly independant from
the previous $a'$ directions and one hyperplane in this new direction.
The obtained configuration ${\mathcal B}'$ is not maximal 
and $N({\mathcal B}')>N({\mathcal B})$, then 
${\mathcal B}$ is not maximal among the ${\mathcal L}$ arrangements
not reaching $N_1$.
\item If all the $d_i$ are $0$ or $q-1$, namely
${\mathcal B}$ is contituted by $a'$ complete blocks
containing $q-1$ hyperplanes. As ${\mathcal B}$ is not maximal,
$d'<d$ holds. Let us add a new hyperplane in a new direction
linearly independant from the $a'$ previous directions. 
As ${\mathcal B}$ is not a $T_4$ configuration, the obtained configuration
${\mathcal B}'$ is not maximal. Moreover $N({\mathcal B}')>N({\mathcal B})$, then 
${\mathcal B}$ is not maximal among the ${\mathcal L}$ arrangements
not reaching $N_1$.
\end{enumerate}

\section{Main Result for general hypersurfaces}\label{sec:trois}
\subsection{Gr\"obner basis techniques}
We will use a Gr\"obner basis theoretical method similar to
the one used by O. Geil in \cite{Geil} 
to compute the second weight
of the generalized Reed-Muller code $RM_q(d,n)$
($q \geq 3$ and $q \leq d \leq (n-1)(q-1)$). 
For the convenience of
the reader we recall some general definitions and
results on Gr\"obner basis
which can be found in \cite{cls}.
We repeat the relevant material from \cite{Geil} and \cite{Geil2},
where the details can be found.

Let ${\mathcal M}$ the set of monomials of $\F_q[X_1,\ldots,X_n]$
$$M(X_1,X_2,\ldots,X_n)=\prod_{i=1}^n X_i^{\alpha_i},$$
where $\alpha_i \in \N$.
Let $\prec$ be a monomial ordering on ${\mathcal M}$.
If $f\in \F_q[X_1,\ldots,X_n]$, we will denote by ${\rm lm}(f)$
its leading monomial and by ${\rm lt}(f)$ its leading term.
We will denote by $\lcm(f,g)$ the
low common multiple of $f$ and $g$.
If ${\rm lm}(f)=\prod_{i=1}^n X_i^{\alpha_i}$,
the multidegree of $f$, denoted by ${\rm multideg}(f)$, is
$(\alpha_1,\cdots,\alpha_n)$.

The first main tool is the division algorithm 
of a polynomial $f \in \F_q[X_1,\ldots,X_n]$
by an ordered set $(f_1,\cdots,f_s)$ of polynomials.
Using this algorithm, $f$ can be written
$$f=a_1f_1+\ldots+a_sf_s +r,$$
where $r \in \F_q[X_1,\ldots,X_n]$ and either $r=0$
or $r$ is a linear combination, with coefficients in $\F_q$,
of monomials, none of which is divisible by any of
$lt(f_1), \ldots, lt(f_s)$. Moreover if $a_if_i \neq 0$,
then we have ${\rm multideg}(f) \preceq {\rm multideg}(a_if_i)$.
Note that the result depends on the monomial ordering and on
the ordering of the $s$-tuple of polynomials $(f_1,\cdots,f_s)$.

\begin{defi}
 Let $\prec$ be a monomial ordering. A finite subset
$\{ g_1,\ldots,g_s\}$ of an ideal $I$ is said to be a
Gr\"obner basis if 
$$\langle {\rm lt}(g_1),\ldots, {\rm lt}(g_s)\rangle=
\langle {\rm lt}(I) \rangle.$$
\end{defi}

The Buchberger's algorithm provides a way to decide if
a basis $\{g_1,\ldots,g_s \}$ is a Gr\"obner basis or not.
It uses the following notion of $S$-polynomial. 

\begin{defi}\label{spolynomial}
 Let $f,g \in \F_q[X_1,\ldots,X_n]$ be two nonzero polynomials.
The $S$-polynomial of $f$ and $g$ is
$$S(f,g)=\frac{\lcm\left(\strut{\rm lm(f)},{\rm lm(g)}\right)}
{{\rm lt(f)}}f - 
\frac{\lcm\left(\strut{\rm lm(f)},{\rm lm(g)}\right)}
{{\rm lt(g)}}g.
$$
\end{defi}

\begin{theo}[Bruchberber's algorithm]\label{bruchberger}
A set $\{g_1,\ldots,g_s\}$
is a Gr\"obner basis for the ideal $\langle g_1,\ldots,g_s \rangle$
if and only if for all pair $i \neq j$
the remainder on division of $S(g_i,g_j)$ by 
$\{g_1,\ldots,g_s\}$ listed in some order is zero.
\end{theo}

\begin{rema}\label{coprime}
The previous algorithm can be simplified by the following remark:
if ${\rm lm}(g_i)$ and ${\rm lm}(g_j)$ are relatively prime, 
then the remainder
on division of $S(g_i,g_j)$ by 
$\{g_1,\ldots,g_s\}$ listed in some order is zero.
\end{rema}

\begin{defi}
Let $I$ be an ideal of $\F_q[X_1,\ldots,X_n]$.
The footprint of $I$ is
$$\Delta(I)=\{M \in {\mathcal M}~|$$
$$M \hbox{ is not the leading monomial of any polynomial in } I\}.$$ 
\end{defi}
We will use the following result which can be found in \cite{Geil2}:
\begin{theo}
Let us consider the following ideal $I$ of $\F_q[X_1,\ldots,X_n]$:
$$I=\langle F_1, \ldots, F_k,X_1^q-X_1, \ldots, X_n^q-X_n \rangle.$$
Then the footprint $\Delta(I)$ is finite and
$$\# \Delta(I)=\# V_q(I)$$
where $V_q(I)$ is the set of the $\F_q$-rational points of the
variety defined by the ideal $I$.
\end{theo}

If we know a Gr\"obner basis of the ideal $I$,
the footprint is easy to determine.

\begin{theo}\label{gbfp}
 Let $I$ be an ideal and $\{g_1,\ldots,g_s\}$ a Gr\"obner
basis of $I$. Let $J$ be the ideal 
$\langle {\rm lm}(g_1),\ldots,{\rm lm}(g_s)\rangle$.
Then
$$\Delta(I)=\Delta(J).$$
\end{theo}

In the following, we will restrict 
$\prec$ to be the graded lexicographic ordering on ${\mathcal M}$ 
defined by
$$\prod_{i=1}^n X_i^{\alpha_i} \prec \prod_{i=1}^n X_i^{\beta_i} $$
if $(\alpha_1, \ldots, \alpha_n) \neq (\beta_1, \ldots, \beta_n)$
and either $\sum_{i=1}^n \alpha_i < \sum_{i=1}^n \beta_i$
holds or $\sum_{i=1}^n \alpha_i = \sum_{i=1}^n \beta_i$
with the first non-zero  entry of 
$(\beta_1-\alpha_1,\ldots,\beta_n-\alpha_n)$ being positive holds.

\subsection{The second weight}
\begin{theo}
For $n\geq 3$, $q\geq 3$ and $q-1< d \leq (n-1) (q-1)$
the second weight $W_2$ of the generalized Reed-Muller code ${\rm RM}_q(d,n)$
satisfies
\begin{enumerate}
 \item if  $1 \leq a \leq n-1$ and $b=0$ then
  $$W_2=W'_2= 2 q^{n-a-1}(q-1);$$
 \item if  $1 \leq a < n-1$ and $b=1$ then
   \begin{enumerate}
      \item if $a<n-2$ then 
$$q^{n-a}-q^{n-a-1} +q^{n-a-2}-q^{n-a-3} \leq W_2 \leq  q^{n-a}=W'_2;$$
      \item if $a=n-2$ then
$$q^{2}-2 \leq W_2 \leq  q^{2}=W'_2;$$
   \end{enumerate} 
 \item if  $1 \leq a < n-1$ and $2\leq b <q-1$ then
      $$W_2=W'_2=q^{n-a-2} (q-1) (q-b+1);$$
\end{enumerate}
\end{theo}
\noindent {\bf Proof.} 
Let $F(X_1,X_2,\ldots, X_n)$ be a reduced polynomial of degree $d$,
${\rm lm}(F)=X_1^{u_1}X_2^{u_2}\ldots X_n^{u_n}$ 
its leading monomial.
We suppose that the variables $X_i$ are numbered in such a way that
$u_1 \geq u_2 \ldots \geq u_n$.
Let us consider the ideals 
$$I=\langle F,X_1^q-X_1,\ldots,X_n^q-X_n \rangle,$$
and 
$$J=\langle X_1^{u_1}X_2^{u_2}\ldots X_n^{u_n},
X_1^q,\ldots,X_n^q \rangle.$$
Using the footprint of $I$ and $J$ we get
$$\# \Delta(I) \leq \# \Delta(J)=q^n - \prod_{i=1}^n (q-u_i).$$
We remark that this last value is the number of points of a hyperplane arrangement
${\mathcal A}$ which is in ${\mathcal L}$.
Then, if 
$(u_1,u_2, \ldots, u_n) \neq (q-1,q-1, \ldots, q-1,b,0
\ldots,0),$
the arrangement ${\mathcal A}$ is not maximal and consequently
$$\# \Delta(I) \leq \# \Delta(J) \leq N'_2.$$

If $(u_1,u_2, \ldots, u_n) = (q-1,q-1, \ldots, q-1,b,0
\ldots,0),$
let us compute for each $1 \leq i \leq a+1$ (or  $1 \leq i \leq a$
if $b=0$) 
$$H_i(X_1,X_2,\ldots, X_n) =\frac{\lcm({\rm lm}(F),X_i^q)}{X_i^q}(X_i^q-X_i) - 
\frac{\lcm({\rm lm}(F),X_i^q)}{{\rm lt}(F)}F,$$
$$H_i(X_1,X_2,\ldots, X_n)=-X_1^{u_1}\ldots X_{i-1}^{u_{i-1}}
X_i X_{i+1}^{u_{i+1}}\ldots X_n^{u_n} - X_i^{q-u_i} G$$
where $G=F-{\rm lt}(F)$.
Then, let us set $R_i$
the remainder of the division of $H_i$ by $( F,X_1^q-X_1,\ldots, X_n^q-X_n)$.
By Bruchberger's algorithm \ref{bruchberger} and Remark
\ref{coprime} ($X_i^q$ and $X_j^q$ are relatively
prime if $i \neq j$) if all the $R_i$ are null, then $\{F,X_1^q-X_1,\ldots,X_n^q-X_n\}$
is a Gr\"obner basis.  Hence by Theorem \ref{gbfp}
$$\# \Delta(I) = \# \Delta(J)=q^n - \prod_{i=1}^n (q-u_i)=q^n-(q-b)q^{n-a-1}.$$
We conclude that in this case the hypersurface defined 
by $F$ is maximal.

If one of the $R_i$ is not zero,
let us consider 
$$M={\rm lm}(R_i)=X_1^{\alpha_1} \ldots X_n^{\alpha_n}.$$
If the index $i$ is such that $1 \leq i \leq a$ we can suppose that $i=1$.
In this case we have $X_1^{q-u_i}=X_1$. Then we have
the following constraints on the exponents 
$(\alpha_1,\ldots,\alpha_n)$:
\begin{enumerate} 
\item $\sum_{i=1}^n {\alpha_i} \leq d+1$,
\item $0 \leq \alpha_i \leq q-1$,
\item $X_1^{q-1} \ldots X_a^{q-1} X_{a+1}^b \hbox{ does not divide } M$.
\end{enumerate}
If $i=a+1$ then $X_i^{q-u_i}=X_{a+1}^{q-b}$. In this case
we have
the following constraints on the exponents 
$(\alpha_1,\ldots,\alpha_n)$:
\begin{enumerate} 
\item $\sum_{i=1}^n {\alpha_i} \leq d+q-b$,
\item $0 \leq \alpha_i \leq q-1$,
\item $X_1^{q-1} \ldots X_a^{q-1} X_{a+1}^b \hbox{ does not divide } M$.
\end{enumerate}

\begin{rema} Let us remark that if $b=\alpha_{a+1}=0$
the first constraint on the $\alpha_i$ is 
always $\sum_{i=1}^n {\alpha_i} \leq d+1$.
\end{rema}

Now we have
$$I=\langle F,R_i,X_1^q-X_1,\ldots,X_n^q-X_n \rangle,$$
so, if we set 
$$J_1=\langle X_1^{q-1} \ldots X_a^{q-1} X_{a+1}^b
 ,M,X_1^q,\ldots,X_n^q \rangle,$$
we get
$$\# \Delta(I) \leq \# \Delta(J_1).$$
Let us consider
$$A_1=\{\beta=(\beta_1,\ldots,\beta_n)~|~\beta_1=q-1, \ldots=\beta_a=q-1,
\beta_{a+1}\geq b,$$
$$0 \leq \beta_{a+2}\leq q-1, \ldots ,
0 \leq \beta_n \leq q-1\}$$
$$A_2=\{\beta=(\beta_1,\ldots,\beta_n)~|~\alpha_1 \leq \beta_1 \leq q-1,\ldots, 
\alpha_n \leq \beta_n \leq q-1\}$$
and
$$A_1\cap A_2=\{\beta=(\beta_1,\ldots,\beta_n)~|~\beta_1=\ldots=\beta_a=q-1,
\beta_{a+1}\geq \gamma,$$
$$\alpha_{a+2} \leq \beta_{a+2} \leq q-1,\ldots, \alpha_n \leq \beta_n \leq q-1\}$$
where $\gamma=max(b,\alpha_{a+1})$.
Then
$$N(F) \leq \# \Delta(J_1)=q^n-\#A_1-\#A_2+\#A_1\cap A_2,$$
$$W(F) \geq \#A_1+\#A_2-\#A_1\cap A_2,$$
$$W(F) \geq (q-b) q^{n-a-1} + \prod_{i=1}^n (q-\alpha_i)
-(q-\gamma)\prod_{i=a+2}^n (q-\alpha_i).$$
The following lemma \ref{lem:ineq} is exactly
what we need to compute the minimum $\mu$ of $\#A_2-\#A_1\cap A_2$.
Then, a lower bound of $W_2$ is $\mu + (q-b) q^{n-a-1}$.
In most cases, namely when
$b \neq 1$, this lower bound is effectively reached by
a hyperplane arrangement and we have $W_2=W'_2$.
 
\begin{lemm}\label{lem:ineq}
Let $q$, $n$, $d$ be integers such
that $q\geq 3$, $n \geq 3$, $q \leq d \leq (n-1)(q-1)$.
We denote by $a$ and $b$ the quotient and the remainder on
division of $d$ by $q-1$, namely
$d=a(q-1)+b$ where $0\leq b <q-1$.

\smallskip

We denote by $V$ the set of the finite sequences of integers
$\alpha=(\alpha_1,\ldots,\alpha_n),$
of length $n$, such that
\begin{enumerate}
 \item for $i=1,\ldots,n$
we have $0 \leq \alpha_i \leq q-1$;
 \item $\sum_{i=1}^n \alpha_i \leq K$
where $K=d+1$ if $b=0$ and $K=d+q-b$ if $b>0$;
 \item if $\alpha_1=\alpha_2=\ldots=\alpha_a=q-1$,
then $\alpha_{a+1}<b$.
\end{enumerate}
Let us set $\gamma=\max(\alpha_{a+1},b)$.

\smallskip

Then, the following holds:
\begin{equation}\label{eq:main}
\min_{\alpha \in V}\left\{\prod_{i=1}^n (q-\alpha_i) -
(q-\gamma)\prod_{i=a+2}^n (q-\alpha_i)\right\} = \mu,
\end{equation}
where 
$$\mu=
\left \{ 
 \begin{array}{ccl}
   (q-2)q^{n-a-1} & \hbox{ if } & b=0\\
    (q-1)q^{n-a-3} & \hbox{ if } & b=1, a<n-2\\
    (q-2)q^{n-a-2} & \hbox{ if } & b=1, a=n-2\\
    (b-1)q^{n-a-2} & \hbox{ if } & 2 \leq b < q-1
 \end{array}
\right .
.$$ 
\end{lemm}

\section{Open questions}\label{sec:quatre}
Now we know the second weight of a Generalized
Reed-Muller code, in almost any case. It remains to
determine the exact value of this second weight when
$d=a(q-1)+1$. For these particular values
we have just proved that
$$q^{n-a}-q^{n-a-1} +q^{n-a-2}-q^{n-a-3} \leq W_2 \leq  q^{n-a}=W'_2
\quad \hbox{ if } a<n-2,$$
and that
$$q^{n-a}-2q^{n-a-2} \leq W_2 \leq  q^{n-a} \quad \hbox{ if } a=n-2.$$
It would be very surprising to find a non-maximal hypersurface
of degree $d=a(q-1)+1$
with strictly more than $q^n-q^{n-a}$ points.
Then we can ask the following questions:
\begin{enumerate}
 \item When $d=a(q-1)+1$, what is the exact value of $W_2$?
 \item When $d=a(q-1)+1$, what is the maximal number of points
of a non-maximal hypersurface 
of degree $d$ given by 
unions of hyperplanes?
(in this paper we have proved that the maximum
number of points for a
hyperplane configuration in ${\mathcal L}$ 
is $q^n -q^{n-a}$).
\end{enumerate}

We have not determined in the paper which are the codewords
reaching the second weight. In our opinion, these codewords
are hyperplanes arrangements. But this is not proved.
However, we can deduce from the results obtained in \cite{RR} on the number
of points of irreducible but not absolutely irreducible
hypersurfaces that such a hypersurface cannot reach the second weight.
In fact a simple computation shows that the number of points of
such a hypersurface is strictly less than the maximum number of points
of a non-maximal hypersurface in ${\mathcal L}$ 
(namely the number called $N'_2=q^n-W'_2$) and {\it a fortiori}
cannot reach the second weight.

\appendix
\section{Proof of lemma \ref{lem:ineq}}
\subsection{Preliminary remarks}
Let us set
$$P_1=\prod_{i=1}^n (q-\alpha_i),$$
$$P_2=(q-\gamma)\prod_{i=a+2}^n (q-\alpha_i).$$
Hence we have to study the minimum value of $P_1-P_2$.
Note that in the particular case $d=(n-1)(q-1)$
the value of $a$ is $n-1$ and $P_2=(q-\gamma)$.

\medskip

\begin{lemm}\label{fct:1}
If we permute the  first $a$ elements $\alpha_i$
we don't change the value of $P_1-P_2$. 
When $\alpha_{a+1} < b$, 
if we permute the last $n-a-1$ elements
we don't change the value of $P_1-P_2$.
When $\alpha_{a+1} \geq b$, namely when $\gamma=\alpha_{a+1}$, 
if we permute $\alpha_{a+1}$ with one of the  
last $n-a-1$ elements $\alpha_i$ such that $\alpha_i\geq b$
we don't change the value of $P_1-P_2$.  
\end{lemm}
\noindent {\bf Proof.} This can be seen directly on
the formulas giving $P_1$ and $P_2$.

\smallskip

Then, from now on, we will suppose that the sequences $\alpha$
are such that
$$\alpha_1 \geq \alpha_2 \geq \cdots \geq \alpha_{a-1} \geq \alpha_a,$$
$$\alpha_{a+2} \geq \alpha_{a+3} \geq \cdots \geq \alpha_{n}$$
if $\alpha_{a+1}<b$, and that
$$\alpha_1 \geq \alpha_2 \geq \cdots \geq \alpha_{a-1} \geq \alpha_a,$$
$$\alpha_{a+1} \geq \alpha_{a+2} \geq \cdots \geq \alpha_{n}$$
if $\alpha_{a+1}\geq b$.
In particular, when we transform a sequence,
we always reorder the new obtained sequence in this way.
 
\begin{lemm}\label{fct:2}
If we replace $\alpha_i$ by $\alpha_i+1$
and if the new sequence is in $V$, then the new
$P_1-P_2$ is lower than the old one. 
\end{lemm}
\noindent {\bf Proof.} When $i \leq a$ the value of $P_1$
decreases, the value of $P_2$ is not modified.
When $i=a+1$ and $\alpha_{a+1}<b$, $P_1$
decreases, the value of $P_2$ is not modified.
When $i \geq a+1$ and $\alpha_{a+1} \geq b$,
$P_2$ and $P_1$ decreases,
then we must examine more precisely the behaviour of $P_1-P_2$.
The difference between the old value of $P_1-P_2$
and the new one is
 \renewcommand{\arraystretch}{0.5}
$$\prod_{\begin{array}{c} j=1\\j\neq i \end{array}}^n
\left ( q-\alpha_j\right) 
-  \prod_{\begin{array}{c} j=a+1\\j\neq i \end{array}}^n 
\left ( q-\alpha_j\right)=$$
$$\left ( 
\prod_{j=1}^a
\left ( q-\alpha_j\right)-1 \right )
\prod_{\begin{array}{c} j=a+1\\j\neq i \end{array}}^n 
\left ( q-\alpha_j\right).$$
 \renewcommand{\arraystretch}{1}
But, as $\alpha_{a+1} \geq b$, $\alpha_a \leq q-2$ and
then 
$$\prod_{j=1}^a
\left ( q-\alpha_j\right) \geq 2.$$
We conclude that the new value is lower than the old one.
It remains to study the case where $\alpha_{a+1}<b$ and $i>a+1$.
The difference between the old value of $P_1-P_2$
and the new one is now
\renewcommand{\arraystretch}{0.5}
$$\left (
\prod_{j=1}^{a+1}
\left ( q-\alpha_j\right)-(q-b) \right )
\prod_{\begin{array}{c} j=a+2\\j\neq i \end{array}}^n 
\left ( q-\alpha_j\right).$$
\renewcommand{\arraystretch}{1}
But as $(q-\alpha_{a+1}) > (q-b)$ we conclude
that the new value of $P_1-P_2$ is lower than the old one.

\begin{lemm}\label{fct:3}
 The minimum in the equation (\ref{eq:main}) is reached
for sequences $\alpha$ such that $\sum_{i=1}^n \alpha_i=K$.
\end{lemm}
\noindent {\bf Proof.} It is sufficient to prove that if 
$\sum_{i=1}^n \alpha_i<K$ it is possible to add $1$
to a well choosen $\alpha_i$ (and then increase the sum),
and obtain a new sequence in $V$ for which the new $P_1-P_2$
is lower than the old one.
Suppose that $\sum_{i=1}^n \alpha_i<K$. 

\medskip

\noindent 1) Suppose that $\alpha_{a+1}\geq b$.
 
\noindent \quad a) If  
$(\alpha_1,\ldots,\alpha_a)\neq (q-1,\ldots,q-1,q-2)$
then there exists a $i$ such that $1 \leq i \leq a$
which does not reach its maximal value. Then
we can replace $\alpha_i$ by $\alpha_i+1$. 
By Lemma \ref{fct:2} 
we conclude that the new $P_1-P_2$ is lower than the old one.

\noindent \quad b) Now suppose
that $(\alpha_1,\ldots,\alpha_a)= (q-1,\ldots,q-1,q-2)$.

\noindent \quad \quad $\alpha$) If $b=0$ then $K=d+1=a(q-1)+1$.
But the sum of the first $a$ elements is $a(q-1)-1$.
So that $\alpha_{a+1}$ is at most $1$ (this term exists
because $a\leq n-1$). Then we can replace $\alpha_{a+1}$
by $\alpha_{a+1}+1$ because $q\geq 3$. By Lemma \ref{fct:2}
we conclude that the new $P_1-P_2$ is lower than the old one.

\noindent \quad \quad $\beta$) If $b \geq 1$ 
then $K=d+q-b=a(q-1)+q$. In this case we know that
$a \leq n-2$. We have $\alpha_{a+1}+\alpha_{a+2}+\ldots \alpha_n \leq q$
then if $\alpha_{a+1}<q-1$ we can add $1$ to this term,
if $\alpha_{a+1}=q-1$ then $\alpha_{a+2} \leq 1$ and
because $q \geq 3$ it is possible to add $1$ to to this term.
By Lemma \ref{fct:2} 
we conclude that the new $P_1-P_2$ is lower than the old one.

\medskip 

\noindent 2) Suppose that $\alpha_{a+1} < b$. Then
by Lemma \ref{fct:2} if we replace $\alpha_{a+1}$ by
$\alpha_{a+1}+1$, we obtain a new $P_1-P_2$ lower than
the old one. 

\medskip

\noindent From now on we will suppose that 
$\alpha$ is such that 
$$\sum_{i=1}^n \alpha_i=K$$.
\begin{lemm}\label{fct:4}
Let $1\leq i \leq a$ and $a+1 \leq j \leq n$ and
suppose that $\alpha_j>\alpha_i$. If we permute these two elements,
and if we obtain a sequence which is in $V$, then for
the new sequence the value of $P_1-P_2$ is lower or
equal to the old one.
\end{lemm}
\noindent {\bf Proof.} Indeed $P_1$ does not change, and $P_2$ increases
(if $j>a+1$ or if $j=a+1$ and $\alpha_j > b$) or does not
change (if $j=a+1$ and $\alpha_j \leq b$).

\begin{lemm}\label{fct:5}
Suppose that $1\leq \alpha_i \leq \alpha_j \leq q-2$
and that we are in one of the following cases:
\begin{enumerate}
\item $1 \leq j < i \leq a$;
\item $a+2 \leq j < i \leq n$;
\item $\alpha_{a+1} \geq b$ and $a+1 \leq j < i \leq n$;
\item $1 \leq j \leq a$ and $a+2 \leq i \leq n$.
\end{enumerate}
Let us replace 
$\alpha_i$ by $\alpha_i-1$ and
$\alpha_j$ by $\alpha_j+1$. If the new
sequence is in $V$, the new value of
$P_1-P_2$ is lower than the old one. 
\end{lemm}
\noindent {\bf Proof.} 1) Case $1 \leq j < i \leq a$. 
The difference between the old value of $P_1-P_2$ and 
the new value is 
$$(\alpha_j-\alpha_i+1) \prod_{k \neq i,j}(q-k) > 0.$$

\noindent 2) Case $a+2 \leq j < i \leq n$.
The difference between the the old value of $P_1-P_2$ and 
the new value is 
\renewcommand{\arraystretch}{0.5}
$$(\alpha_j-\alpha_i+1) 
\left (
\prod_{k=1}^{a+1}
\left ( q-\alpha_k\right)-(q-\gamma) \right )
\prod_{\begin{array}{c} k=a+2\\k\neq i,j \end{array}}^n 
\left ( q-\alpha_k\right) >0 .$$
To verify that the previous expression is $>0$
note that if $\alpha_{a+1} <b $ 
then $\gamma=b$ and
$$\prod_{k=1}^{a+1}
\left ( q-\alpha_k\right) \geq (q-\alpha_{a+1}).$$
Hence 
$$\prod_{k=1}^{a+1}
\left ( q-\alpha_k\right)-(q-b)\geq (b-\alpha_{a+1})>0.$$
If $\alpha_{a+1} \geq b$ then $\gamma=\alpha_{a+1}$ and 
$\alpha_a \leq q-2$. Then
$$\prod_{k=1}^{a+1}
\left ( q-\alpha_k\right)-(q-\gamma) \geq 2(q-\alpha_{a+1}) -
(q-\alpha_{a+1})=(q-\alpha_{a+1}) >0.$$
\renewcommand{\arraystretch}{1}
 
\noindent 3) Case $\alpha_{a+1} \geq b$ and $a+1 \leq j < i \leq n$.
The formula of the difference
between the the old value of $P_1-P_2$ and 
the new value is similar 
\renewcommand{\arraystretch}{0.5}
$$(\alpha_j-\alpha_i+1) 
\left (
\prod_{k=1}^{a}
\left ( q-\alpha_k\right)-1 \right )
\prod_{\begin{array}{c} k=a+1\\k\neq i,j \end{array}}^n 
\left ( q-\alpha_k\right) > 0.$$
To verify that the previous expression is $>0$ we have just to remark
that $\alpha_a \leq q-2$, then 
$\prod_{k=1}^{a}\left ( q-\alpha_k\right)\geq 2$.
\renewcommand{\arraystretch}{1}

\noindent 4) Case $1 \leq j \leq a$ and $a+2 \leq i \leq n$.
A simple computation shows that the difference between
the old value of $P_1-P2$ and the new value
is
$$(\alpha_j-\alpha_i+1)\prod_{\begin{array}{c} k=1\\k\neq i,j 
\end{array}}^n 
\left ( q-\alpha_k\right)\quad 
+\quad (q-\gamma) \prod_{\begin{array}{c} 
 k=a+2\\\ k\neq i \end{array}}^n 
\left ( q-\alpha_k\right) >0
$$

\subsection{The head of a best sequence}
We give here the form of the  first $a$ terms
of a sequence $\alpha$ for which $P_1-P_2$
is minimum. We prove that $\alpha$ can be 
choosen such that one of the two following conditions
holds:
\begin{enumerate}
 \item $(\alpha_1,\ldots \alpha_{a-1}, \alpha_a)=
(q-1,\ldots,q-1,q-2)$ and $\alpha_{a+1} \geq b$;
\item $(\alpha_1,\ldots , \alpha_a)=(q-1,\ldots,q-1)$
and $\alpha_{a+1} < b$;
\end{enumerate}

\medskip

\noindent 1) Let us suppose first that there exists 
a sequence $\alpha$ such that $\alpha_{a+1}<b$ and
for which 
$P_1-P_2$ is minimum.
We will prove that for such a sequences the  first $a$ terms 
can be set to $q-1$.
Suppose that there exists a $j\leq a$ such that $\alpha_j<q-1$.
We have
$$\sum_{i=1}^{a+1} \alpha_i <a(q-1)+b=d,$$
then $\alpha_{a+2}>0$. If $\alpha_{a+2}>\alpha_j$ by Lemma \ref{fct:4}
we can permute the two terms to obtain a sequence which have a lower
or equal $P_1-P_2$. If $\alpha_{a+2}\leq \alpha_j$, by Lemma \ref{fct:5}
the sequence obtained by replacing $\alpha_{a+2}$ by $\alpha_{a+2}-1$
and $\alpha_j$ by $\alpha_j+1$ has a lower $P_1-P_2$.
So we have proved that we can increase the value of $\alpha_j$
and obtain a lower or equal $P_1-P_2$.

\medskip

\noindent 2) Let us suppose now that 
there exists 
a sequence $\alpha$, such that $\alpha_{a+1}\geq b$, 
for which 
$P_1-P_2$ is minimum.
We will prove that for such a sequence the 
 first $a-1$ terms 
can be set to $q-1$ and $\alpha_{a}$ can be set to $q-2$.
Suppose that there exists a $j\leq a$ such that $\alpha_j<q-1$
if $j<a$ or 
or that $\alpha_{j}<q-2$ if $j=a$.

\noindent \quad a) If $b=0$ then $K=d+1=a(q-1)+1$.
But $\sum_{i=1}^{a} \alpha_i < a(q-1)-1$. Then $\alpha_{a+1}>0$.
If $\alpha_{a+1}> \alpha_j$ by Lemma \ref{fct:4}
we can permute the two terms to obtain a sequence which have a lower
or equal $P_1-P_2$. If $\alpha_{a+1}\leq \alpha_j$, by Lemma \ref{fct:5}
the sequence obtained by replacing $\alpha_{a+1}$ by 
$\alpha_{a+1}-1$
and $\alpha_j$ by $\alpha_j+1$ has a lower $P_1-P_2$.
So we have proved that we can increase the value of $\alpha_j$
and obtain a lower or equal $P_1-P_2$.

\noindent \quad b) If $b>0$ then $K=d+q-b=a(q-1)+q$.
But $\sum_{i=1}^{a} \alpha_i < a(q-1)-1+$
and then $\sum_{i=1}^{a+1} \alpha_i < a(q-1)+q-2+$.
Hence $\alpha_{a+2}>0$. With the same method than 
in the previous part 1) we prove that
we can increase the value of $\alpha_j$
and obtain a lower or equal $P_1-P_2$.

\subsection{The tail of a best sequence}
We give here the form of the terms $\alpha_i$
for $i \geq a+1$ 
of a sequence $\alpha$ for which $P_1-P_2$
is minimum, assuming that the head is
as in the previous subsection.

\medskip

\noindent 1) Let us suppose first that there exists 
a sequence $\alpha$ such that $\alpha_{a+1}<b$ and
for which 
$P_1-P_2$ is minimum. 
We have seen in the previous subsection 
that we can suppose that the  first $a$ terms are 
$q-1$. We know that $K=a(q-1)+q$. Then 
$\sum_{i=a+2}^{n} \alpha_i=q-\alpha_{a+1}$
using Lemma \ref{fct:5} we can pack the terms $\alpha_i$ for
$i \geq a+2$ in such a way that 

\noindent \quad a)
if $\alpha_{a+1}=0$
then $\alpha_{a+2}=q-1$, $\alpha_{a+3}=1$ and $\alpha_i=0$
for $i>a+3$;

\noindent \quad b) if $b>\alpha_{a+1}\geq 1$ then
$\alpha_{a+2}=q-\alpha_{a+1}$ and 
and $\alpha_i=0$
for $i>a+2$.

\medskip

\noindent 2) Let us suppose now that $\alpha_{a+1}\geq b$.
We We have seen in the previous subsection 
that we can suppose that the $a-1$ first $a-1$ terms are 
$q-1$ and $\alpha_{a}=q-2$.
 
\noindent \quad a) If $b=0$ then $K=a(q-1)+1$. Then
by Lemma \ref{fct:5} we can pack the terms $\alpha_i$ for
$i \geq a+1$ in such a way that $\alpha_{a+1}=2$
and $\alpha_{i}=0$ for $i>a+1$.

\noindent \quad b) If $b>0$ then $K=a(q-1)+q$. Then
by Lemma \ref{fct:5} we can pack the terms $\alpha_i$ for
$i \geq a+1$ in such a way that $\alpha_{a+1}=q-1$, $\alpha_a+2=2$
and $\alpha_{i}=0$ for $i>a+2$.

\subsection{The minimum value of $P_1-P_2$}
~

\noindent 1) Case b=0. Then $K=a(q-1)+1$. The previous results
give directly a sequence for which $P_1-P_2$ is minimum:
$$\alpha_1=\cdots =\alpha_{a-1}= q-1,$$
$$\alpha_a=q-2, \quad \alpha_{a+1}=2, \quad \alpha_{a+2}=\cdots\alpha_n=0.$$
For this sequence we have
$$P_1=2(q-2)q^{n-a-1} \quad P_2= (q-2)q^{n-a-1},$$
then the minimum value of $P_1-P_2$ is
$$\mu = (q-2)q^{n-a-1}.$$

\medskip

\noindent 2) Case b=1. Then $K=a(q-1)+q$ and $a \leq n-2$. 

\noindent \quad a) Let us test first the assumption $\alpha_{a+1}=0$. 
The previous results
give directly a sequence reaching the minimum of $P_1-P_2$
under this assumption:
$$\alpha_1=\cdots =\alpha_{a}= q-1,$$ 
$$\alpha_{a+1}=0,\quad \alpha_{a+2}=q-1, \quad \alpha_{a+3}=1, \quad
\alpha_{a+4}=\cdots\alpha_n=0.$$
We remark that if $a=n-2$ this case cannot occur
because there is not enough room to contain all
the $\alpha_i$.
For this sequence we have
$$P_1=q(q-1)q^{n-a-3},\quad P_2=(q-1)(q-1)q^{n-a-3},$$
so that the minimum of $P_1-P_2$ under this assumption
is 
$$\mu_1=(q-1)q^{n-a-3}.$$

\noindent \quad b) Now let us test the assumption  
$\alpha_{a+1}\geq b=1$. The previous results 
give directly a sequence reaching the minimum of $P_1-P_2$
under this assumption:

$$\alpha_1=\cdots =\alpha_{a-1}= q-1,$$ 
$$\alpha_{a}=q-2, \quad \alpha_{a+1}=q-1,\quad \alpha_{a+2}=2, \quad 
\alpha_{a+3}=\cdots\alpha_n=0.$$
For this sequence we have
$$P_1=2(q-2)q^{n-a-2},\quad P_2=(q-2)q^{n-a-2},$$
so that the minimum of $P_1-P_2$ under this assumption
is
$$\mu_2=(q-2)q^{n-a-2}.$$

\noindent \quad c) Conclusion on the case $b=1$.
Let us compare $\mu_1$ and $\mu_2$ (when $a<n-2)$):
$$\mu_2-\mu_1=q^{n-a-1}-2q^{n-a-2}-q^{n-a-2}+q^{n-a-3},$$
$$\mu_2-\mu_1=q^{n-a-2}(q-3)+q^{n-a-3}.$$
But $q \geq 3$, then $\mu_2>\mu_1$. 
Hence the minimum value is $\mu_1$.

Let us summarize the obtained result in the case $b=1$:
\begin{itemize}
 \item if $a<n-2$ then $\mu=\mu_1=(q-1)q^{n-a-3}$;
 \item if $a=n-2$ then $\mu=\mu_2=(q-2)q^{n-a-2}=q-2$.
\end{itemize}

\medskip

\noindent 3) Case $2 \leq b < q-1$. Then $K=a(q-1)+q$ and 
$a \leq n-2$. 

\noindent \quad a) Test of the assumption $\alpha_{a+1}<b$.

\noindent \quad \quad $\alpha$) Test of the joint assumption 
$\alpha_{a+1}=0$.
The previous results 
give directly a sequence reaching the minimum of $P_1-P_2$
under this assumption:
$$\alpha_1=\cdots =\alpha_{a}= q-1,$$ 
$$\alpha_{a+1}=0,\quad \alpha_{a+2}=q-1, \quad \alpha_{a+3}=1, \quad
\alpha_{a+4}=\cdots\alpha_n=0.$$
This case cannot occur if $a=n-2$. 
For this sequence we have
$$P_1=q(q-1)q^{n-a-3},\quad P_2=(q-b)(q-1)q^{n-a-3}.$$
Then
the minimum reached by $P_1-P_2$ under this assumption is
$$\mu_1=b(q-1)q^{n-a-3}.$$

\noindent \quad \quad $\beta$) Test of the joint assumption 
$\alpha_{a+1}\neq 0$. The previous results shows
that a sequence reaching the minimum of $P_1-P_2$
under these assumptions is of the form
$$\alpha_1=\cdots =\alpha_{a}= q-1,$$ 
$$\alpha_{a+1}>0,\quad \alpha_{a+2}=q-\alpha_{a+1}, \quad
\alpha_{a+3}=\cdots\alpha_n=0.$$
For this sequence we have
$$P_1=(q-\alpha_{a+1})\alpha_{a+1}q^{n-a-2},
\quad P_2=(q-b)\alpha_{a+1}q^{n-a-2},$$
then
$$P_1-P_2=(b-\alpha_{a+1})\alpha_{a+1}q^{n-a-2}.$$
The minimum of the quadratic polynomial $(b-\alpha_{a+1})\alpha_{a+1}$
(with $1 \leq \alpha_{a+1} < b \leq q-2$) is reached for
$\alpha_{a+1}=1$ which gives for minimum of $P_1-P_2$
$$\mu_2=(b-1)q^{n-a-2}.$$

\noindent \quad b) Test of the assumption $\alpha_{a+1} \geq b$.
The previous results shows
that a sequence reaching the minimum of $P_1-P_2$
under this assumption is
$$\alpha_1=\cdots =\alpha_{a-1}= q-1,$$ 
$$\alpha_{a}=q-2, \quad \alpha_{a+1}=q-1,\quad \alpha_{a+2}=2, \quad 
\alpha_{a+3}=\cdots\alpha_n=0.$$
For this sequence we have
$$P_1=2(q-2)q^{n-a-2},\quad P_2=(q-2)q^{n-a-2},$$
so that the minimum of $P_1-P_2$ under this assumption
is
$$\mu_3=(q-2)q^{n-a-2}.$$

\noindent \quad c) Conclusion of the case $2 \leq b<q-1$.
The minimum of $P_1-P_2$ is
$$\mu =\min(\mu_1,\mu_2,\mu_3)=\mu_2=(b-1)q^{n-a-2}.$$
Indeed, as $q-1> b\geq 2$, we have $q-2> b-1>0$, which prove that
$\mu_3 > \mu_2$.
To prove that $\mu_1>\mu_2$
let us compute
$$\mu_1-\mu_2=b(q-1)q^{n-a-3}-(b-1)q^{n-a-2} =
q^{n-a-2}-bq^{n-a-3}.$$
But $b<q$, then $\mu_1-\mu_2>0$.

\end{document}